\documentstyle[multicol,prl,aps]{revtex}
\input epsf.tex
\begin{document}                
\draft
\title{Singular Effects of Impurities near the Ferromagnetic Quantum-Critical Point}
\author{H. Maebashi\cite{byline} and K. Miyake}
\address{Department of Physical Science, Graduate School of Engineering Science, Osaka University, 
\\ Toyonaka, Osaka 560-8531, Japan}
\author{C. M. Varma}   
\address{Bell Laboratories, Lucent Technologies, Murray Hill, New Jersey 07974}  
\maketitle
\begin{abstract}                
Systematic theoretical results for the effects of a dilute concentration
of magnetic impurities on the thermodynamic and transport properties
in the region around the quantum critical point of
a ferromagnetic transition are obtained. 
In the quasi-classical regime, 
the dynamical spin fluctuations 
enhance the Kondo temperature. 
This energy scale decreases rapidly 
in the quantum fluctuation regime, where the properties are those of a line of 
critical points of the multichannel Kondo problem with the number of channels 
increasing as the critical point is approached, except at unattainably low temperatures
where a single channel wins out.
\end{abstract}

\begin{multicols}{2}

%
The asymptotic low temperature singularities in the thermodynamic and 
transport properties of many solids appear to be due to impurities \cite{sfl}. 
While several impurity models have quantum-critical points (QCP), 
where such singularities may be expected, 
they require special symmetries unlikely to be present 
in real systems\cite{NB80}. 
An alternate possibility is that the pure system is near a QCP, 
so that singular low energy fluctuations 
are present\cite{Hertz76,Moriya85,Mishra78}. 
We investigate the possibility in this paper that such fluctuations 
drive the effects of a dilute concentration of ordinary impurities 
(not requiring any special symmetries) so that the resulting observable 
properties 
are much more singular than those near pure QCP \cite{hv,morr}. 

For coupling to non-magnetic impurities, 
a general argument on the renormalized impurity scattering was given 
in relation to insulating behavior in marginal Fermi liquids\cite{Varma97}. 
It was recently proposed that critical fluctuations 
enhance potential scattering from non-magnetic 
impurities  so that the residual resistivity increases 
as the QCP is approached\cite{MNO99}. 

Here we investigate the effects of magnetic impurities 
near a ferromagnetic QCP. 
The problem is especially interesting for a number of reasons.
First, it couples the quantum fluctuation or
Kondo effect of magnetic impurities to the
singular quantum fluctuations of the pure system. 
Second, since the transition is at $q=0$ and the order parameter is conserved, 
conservation laws or Ward identities can be used to obtain systematic results 
for the effect of impurities  just as they are available 
for the pure ferromagnetic QCP.
Third, experimental results on extraordinary pure samples of a 
ferromagnet-MnSi\cite{PML95,TFLSJ95}, 
which has a QCP as a function of pressure, 
are in good accord with the theory\cite{Moriya85}. 
A few results on less pure MnSi are available \cite{TFLSJ95}, 
which show interesting deviations from that of 
the purer samples near the QCP. 
We hope our results will serve as an impetus to 
further experimental results. 

Fig.1 presents a schematic phase diagram around a pure ferromagnetic QCP 
at $r$$=$$0$ where $r$ is a "disordering" parameter, pressure for MnSi. 
The crossover between the regimes I and II occurs at 
$T$$\propto$$\xi^{-3}(T)$ where $\xi$ is 
the magnetic correlation length\cite{Hertz76}.
In the pure limit, in the quantum-critical regime, 
the pole in the fluctuation spectra $\chi ({\bf q},\omega)$ has a dispersion 
$\omega$$\sim$$q^3$, so that the dynamical critical exponent 
$z_{\rm d}$$=$$3$. 
Correspondingly, 
$\xi(T)$$\sim$ $T^{-2/3}$ in the region I. 
In the region II, $\xi$$\sim$$r^{-1/2}$ at low $T$. 

The problem of magnetic impurities was looked at long ago by
Larkin and Mel'nikov\cite{LM72}. 
We have obtained some new results. 
In the regime I, 
the net result of enhanced dynamical spin fluctuations 
at low energies and of a decreased high frequency cutoff is to 
enhance the Kondo temperature. 
In the regime II, this enhancement decreases rapidly 
and an ordinary $S$$=$$1/2$ magnetic impurity necessarily scatters in
many angular momentum channels because of the increasing magnetic 
correlation length with the number of channels diverging
as $r$$\to$$0$. The crossover from the multichannel to the single-chan- nel
behavior occurs at unattainably low temperatures. 

We consider the model of $S$$=$$1/2$ magnetic impurities 
coupling  to the host electrons by the Hamiltonian

\begin{equation}
\frac{J}{2 N} \sum_{{\bf k},{\bf k}',\alpha, \beta}
{\mbox{\boldmath $\sigma$}}_{\alpha \beta} \cdot {\bf S} 
c_{{\bf k}'\alpha}^\dagger c_{{\bf k} \beta}^{\phantom \dagger}. 
\label{Hamiltonian}
\end{equation}

\noindent
We assume that the dynamical spin susceptibility of the pure system is given by

\begin{equation}
\chi ({\bf q},\omega) = 
\chi_0 \kappa_0^2
[\kappa^2 + q^2 - {\rm i} \omega /{\mit\Gamma}q]^{-1},
\label{spectra}
\end{equation} 

\noindent
where $\chi_0$ is the magnetic susceptibility of the noninteracting system, 
$\kappa_0$ is of the order of the Fermi wave number $p_{\rm F}$, 
and $\kappa$$\equiv$$\kappa (T)$ is the inverse of $\xi (T)$\cite{Moriya85}. 
Hereafter, $r$ is defined by $r$$\equiv$$\kappa^2 (T$$=$$0)$$/$$4p_{\rm F}^2$.

For coupling to magnetic impurities, the vertex renormalization 
${\Lambda}^{(\sigma)} (p,\varpi) 
$$/$${\Lambda}^{(\sigma)}_0$ is given by 
a formally exact expression in terms of 
the irreducible part $\bar{\Lambda}^{(\sigma)}(p,\varpi)$: 

\begin{equation}
\frac{\Lambda^{(\sigma)}_{\phantom 0}
(p,\varpi)}{{\Lambda}^{(\sigma)}_{0}} 
= 
\left(
1+\alpha\frac{\chi(\varpi)}{\chi_0}
\right)\frac{\bar{\Lambda}^{(\sigma)}(p,\varpi)}{
{\Lambda}^{(\sigma)}_0}, 
\label{exact}
\end{equation}

\noindent
where $\alpha$ is the coupling constant between the host electrons, 
$\alpha$$=$$2U\chi_0$ for the Hubbard interaction. 
We have used the simplified notation, 
$p$$=$$({\bf p}, \varepsilon)$ and $\varpi$$=$$({\bf q},\omega)$.

For fluctuations in which $\omega$$\sim$$q^3$, 
a version of Migdal theorem holds 
in the so-called $q$ limit, $\omega$$\to$$0$, $q$$\to$$0$ with 
$\omega$$/$$v q = 0$ so that the Ward-identity requires \cite{Nozieres97} 

\begin{equation}
\bar{\Lambda}^{(\sigma)}_{\phantom 0}/{\Lambda}^{(\sigma)}_{0}
= [ z \, (m^*/m) \, (\alpha + \chi_0/\chi)]^{-1} 
\simeq 1, 
\label{identity}
\end{equation}

\noindent 
i.e., $z(m^*$$/$$m)$$\simeq$$\alpha^{-1}$$\sim$$1$ 
when $\chi_0$$/$$\chi$$\ll$$1$.
By general considerations, 
$m$$/$$m^*$$=$$z(1$$+$$\partial \Sigma$$/$$\partial \epsilon_{\bf p})$.
So $\partial \Sigma$$/$$\partial \epsilon_{\bf p}$ is not singular while 
$z^{-1}$$=$$1$$+$$\partial \Sigma$$/$$\partial \varepsilon$$
\sim$$|\ln (|\varepsilon|, T)|$. 
Another consequence of Eq.(\ref{identity}) is that 
the imaginary part of the one-interaction irreducible part of $\chi(\varpi)$ 
is given by

\begin{equation}
{\rm Im} {\bar \chi}(\varpi) 
\simeq \frac{\pi}{2} \sum_{\bf k} 
\left[
f(\varepsilon_{{\bf k}'})-f(\varepsilon_{{\bf k}})
\right]
\delta (\omega-\varepsilon_{{\bf k}'}+\varepsilon_{{\bf k}})z^2,
\label{Imchi}
\end{equation}

\noindent 
where ${\bf k}'$$=$${\bf k}$$+$${\bf q}$, 
$\varepsilon_{{\bf k}}$ is the energy of the quasi-particles.

The interaction vertex between the host electrons 
and the pseudofermion field has the form 
$\Gamma$$=$$\Gamma^{(0)}$$+$$\Gamma^{(\sigma)} 
{\mbox{\boldmath $\sigma$}}$$\cdot$${\bf S} $. 
To the leading order in $J$, 
the vertex is renormalized as

\begin{equation}
\Gamma^{(\sigma)} (p,\omega_1;
p+\varpi,\omega_2) 
=\frac{J}{2 N}
\frac{{\Lambda}^{(\sigma)}(p,\varpi)
}{{\Lambda}^{(\sigma)}_0},
\label{rvertex}
\end{equation}

\noindent
where $\omega_1$$-$$\omega_2$$=$$\omega$.
Using Eqs.(\ref{exact}) and (\ref{identity}),   
$\Gamma^{(\sigma)}$ is proportional to $\chi(\varpi)$ 
in the $q$ limit. 
In the three-dimensional system with a spherical Fermi surface, 
it is useful to consider partial-wave components of 
$\Gamma^{(\sigma)}$:

\begin{eqnarray}
& &\Gamma^{(\sigma)} ({\bf p},\varepsilon,\omega_1;
{\bf p}+{\bf q},\varepsilon + \omega,\omega_2) 
= \frac{J}{2 N}
\frac{\alpha \kappa_0^2}{\kappa^2+q^2 - {\rm i} \omega /{\mit\Gamma} q  }
\\
& & =\sum_{l=0}^{\infty}
\frac{2l+1}{2} 
\Gamma_l^{(\sigma)} (\varepsilon,\omega_1;\varepsilon +\omega,\omega_2)
P_l (\cos \theta),
\end{eqnarray}

\noindent
where 
${\bf q}^2$$=$$({\bf p}$$-$${\bf p}')^2$$=$$p^2$$+$$p'^2$$-$$2pp'$$\cos 
\theta $, 
$P_l$ are Legendre functions of the first kind. 
The real part of $\Gamma_l^{(\sigma)}$$(\omega)$$\equiv$ $
\Gamma_l^{(\sigma)}$$(\varepsilon,\omega_1;\varepsilon$$+$$\omega,\omega_2)$ 
has a singular contribution, 
while the imaginary part of $\Gamma_l^{(\sigma)}(\omega)$ 
vanishes when $\omega$$\to$$0$.
By use of three roots, $x_1$,$x_2$ and $x_3$, 
of a cubic equation, 
$x^3$$-$$2 x^2$$+$$x$$-$$(\omega$$/$${\mit\Gamma} \kappa^3)^2$$=$$0$, 
we obtain

\begin{eqnarray}
{\rm Re} \Gamma_l^{(\sigma)} (\omega) = 
\frac{J}{2 N}
\frac{ \alpha \kappa_0^2}{pp'}
\bigg[ & &\frac{x_1(1-x_1)}{(x_3-x_1)(x_1-x_2)} {\cal Q}_l(\zeta_1)
\nonumber
\\
+ & & \,\,\,
2\,\,\,{\rm cyclic}\,\,\,{\rm permutations}
\bigg] ,
\label{ReGamma}
\end{eqnarray}

\noindent
where $\zeta_{i}$$=$$(p^2$$+$$p'^2$$+$$\kappa^2 x_{i} )$$/$$2pp'$ 
for $i$$=$$1$,$2$,$3$ and 
${\cal Q}_l$ are Legendre functions of the second kind. 
Noting that ${\cal Q}_l(\zeta)$ have branch points at $\zeta$$=$$\pm 1$, 
we can evaluate the singular contribution in $\Gamma_l^{(\sigma)}$ as  

\begin{equation}
\Gamma_l^{(\sigma)} (\omega) \simeq 
\frac{J\alpha }{6 N} \frac{\kappa_0^2}{p_{\rm F}^2}
\ln \frac{D_l}{max [
|\omega|, 
{\mit\Gamma} \kappa^3, 
{\mit\Gamma} |p - p'|^3]},
\label{approx}
\end{equation}

\noindent
where 
$D_l$$\equiv$${\mit\Gamma} (2 p_{\rm F} \exp (- \sum_{n=1}^l n^{-1}))^3$ 
serves as an upper cutoff of $|\omega|$, 
${\mit\Gamma} \kappa^3$ or 
${\mit\Gamma} |p - p'|^3$ 
for each $l$. 
For the Kondo problem we will investigate bellow, 
the on-shell scattering 
$\varepsilon_{{\bf p}'}$$=$$\varepsilon_{\bf p}$$+$$\omega$ 
is important so that
$|\omega|$$>$${\mit\Gamma}|p - p'|^3$. 
Hence, we put $p$$=$$p'$$=$$p_{\rm F}$. 

If $|\omega|$ is regarded as $\sim$$T$, 
Eq.(\ref{approx}) leads us to the following facts: 
in the regime II, the $\omega$-dependence of $\Gamma_l^{(\sigma)}$ 
is negligible and the number of effective angular momentum channels 
increases as $l_{\rm max}$$\sim$$2 p_{\rm F} \xi$ because the effective size 
of the impurity scales up as $\xi$\cite{LM72}; 
in the regime I, the $\omega$-dependence of $\Gamma_l^{(\sigma)}$ 
is singular as $\Gamma_l^{(\sigma)}$$\propto$$\ln (D_l/|\omega|)$ 
because the effective exchange interaction acts over the long range 
not only in space but also in time. 

In order to demonstrate carefully 
how the singular $\omega$-dependence is combined with 
perturbative renormalization group (RG) equations 
for the Kondo problem near the QCP, 
first we consider 
the renormalization factor of the pseudofermion propagator 
$Z(\omega)$. 
The leading order correction, $\delta Z(\omega)$, 
with respect to $J$ is given by

\begin{equation}
\delta Z(\omega) = 
-\frac{J^2}{N^2} S(S+1) \sum_{\bf q} \int_{-\infty}^0 
\frac{{\rm d} \omega'}{\pi} 
\frac{ {\rm Im} \chi ( {\bf q} , \omega' ) }{ (\omega + \omega')^2 }. 
\label{deltaZ}
\end{equation}

\noindent
Noting that the imaginary part of $\chi$ is given exactly by 

\begin{equation}
{\rm Im} \chi(\varpi)=
\left| \frac{{\Lambda}^{(\sigma)}(p,\varpi)
}{{\bar {\Lambda}}^{(\sigma)}(p,\varpi)}
\right|^2
{\rm Im} {\bar \chi}(\varpi),
\end{equation}

\noindent 
and using Eqs.(\ref{identity})-(\ref{rvertex}), 
Eq.(\ref{deltaZ}) can be evaluated as  

\begin{equation}
\sim S(S+1) \sum_{l=0}^{\infty} \frac{2 l +1}{2}
\int_0^{D} {\rm d} \varepsilon 
\frac{ |z \Gamma_l^{(\sigma)}(\varepsilon)|^2 
}{ \omega-\varepsilon } 
\frac{m^{*2}}{m^{2{\phantom *}}} \varrho^2 , 
\label{part}
\end{equation}

\noindent
where $\varrho$$=$$N m p_{\rm F}$$/$$2\pi^2$.

From Eq.(\ref{part}), the variation of $Z(\omega_1)$ 
on reducing the bandwidth cutoff
from $D$ to $D$$-$$\delta D$ is obtained immediately.
For $\Gamma_l^{(\sigma)}(\varepsilon_{{\bf p}},\omega_1;
\varepsilon_{{\bf p}'},\omega_2)$, 
the $O(J^2)$ term is

\begin{eqnarray}
-\frac{z}{2}\frac{m^*}{m^{\phantom *}}\varrho
\left[
\Gamma_l^{(\sigma)}(D-\varepsilon_{{\bf p}})
\Gamma_l^{(\sigma)}(\varepsilon_{{\bf p}'}-D)
\frac{
\delta D
}{
\omega_1-D+\varepsilon_{{\bf p}}}
\right. & &
\nonumber
\\
+
\left.
\Gamma_l^{(\sigma)}(-D-\varepsilon_{{\bf p}})
\Gamma_l^{(\sigma)}(\varepsilon_{{\bf p}'}+D)
\frac{
\delta D
}{
\omega_1-D-\varepsilon_{{\bf p}'}}
\right] &. &
\label{2ndv}
\end{eqnarray}

\noindent
The $O(J^3)$ term of $\Gamma_l^{(\sigma)}$  
can be evaluated in a similar way to $\delta Z(\omega)$. 
In Eqs.(\ref{part}) and (\ref{2ndv}), 
$z$ and $m^*$ should be understood as functions of the energy scale $D$, 
respectively. 
The high-energy spectrum around the band edge of the host electrons 
may be approximated to be that of the free electrons so that 
$z$$\sim$$1$ and $m^*$$\sim$$m$ at an early stage of the RG procedure. 
Even at the low energy, where $z$ vanishes on a logarithmic scale,    
the product $z(m^*/m)$ remains of the order of $1$ by virtue of 
Eq.(\ref{identity}).
Therefore the renormalization of the wave function and of the mass 
of the host electron does not affect the resulting scaling equation. 
Neglecting $\varepsilon_{\bf p}$$/$$D$, $\varepsilon_{{\bf p}'}$$/$$D$ 
and $\omega_1$$/$$D$  compared to $1$ 
while keeping the explicit dependence on 
$\omega$$=$$\varepsilon_{{\bf p}'}$$-$$\varepsilon_{{\bf p}}$, 
we obtain the 2-loop scaling equation for the invariant coupling 
$\lambda_l (\omega)$ as

\begin{equation}
\frac{{\rm d} \lambda_l (\omega)}{{\rm d} \ln D} 
= - \left| \lambda_l (D) \right|^2
+ \lambda_l (\omega) \sum_{l'=0}^{\infty} 
\frac{2l'+1}{2} 
\left| \lambda_{l'} (D) \right|^2. 
\label{beta}
\end{equation}

\noindent 
Using Eq.(\ref{approx}), 
the initial bandwidth cutoff in Eq.(\ref{beta}) 
can be put as
$D_0$$=$${\mit\Gamma} (2 p_{\rm F})^3$; 
the bare coupling constant, i.e., 
the value of $\lambda_l (\omega)$ at $D$$=$$D_0$, is given by 
$\lambda_l^{({\rm b})} (\omega)$$=$$( 
\varrho$$/$$\alpha ) \Gamma_{l}^{(\sigma)}(\omega)$ 
and the scaling stops at $D$$=$$T$.

The solution of Eq.(\ref{beta}) has the form, 
$\lambda_l (\omega)$$=$$\lambda_l^{({\rm b})} (\omega) A$ 
$+$$ (J \varrho $$/$$N ) B_l$, 
where $A$ and $B_l$ are functions of $D$ which are independent of $\omega$. 
Correspondingly, 
the form of the effective time-dependent interaction in the momentum space 
divided by $J \varrho $$/$$N$, which is given by

\begin{equation}
\hspace{-0.5cm} 
A \delta_0 {\rm e}^{-\Omega_{\theta} |t-t'|} \sin \frac{\theta}{2}  
+\sum_{l=0}^{l_{\rm max}} (2 l +1) B_l 
\frac{\delta (t-t')}{D} P_l (\cos \theta),\hspace{-0.5cm}
\end{equation}

\noindent
is invariant under the present RG transformation with scaling time 
as $t$$\sim$$1/D$, 
where $\delta_0$$\equiv$$ \kappa_0^2$$/$$4 p_{\rm F}^2$ and 
$\Omega_{\theta}$$\equiv$ $D \sin (\theta$$/$$2) [\sin^2 (\theta$$
/$$2)$$+$$\delta]$ with $\delta$$\equiv$$ \kappa^2$$/$$4 p_{\rm F}^2$. 
The RG flow of $A$ and $B_l$ describes correlation not only between 
angular momentum channels but also between long-time and instantaneous 
components of the effective interaction.

If we consider the scaling equation at the one-loop level 
where the third-order term of Eq.(\ref{beta}) is neglected, 
we can make a rough estimation of the Kondo temperature $T_{\rm K}$, 
which is associated with the breakdown of the perturbation theory, 
near the QCP. 
By use of Eq.(\ref{approx}) for $\lambda_l^{({\rm b})} (\omega)$, 
the one-loop scaling equation can be solved analytically. 
The result depends on
whether ${\mit\Gamma} \kappa^3(T)$ is smaller than $T$ or not 
in accordance with the region I or II as 

\[
\lambda_{l}(0) = \left\{
\begin{array}{ll}
{\tilde J}
\ln (T/{\mit\Gamma} \kappa^3) 
+\sqrt{{\tilde J}} 
\tan \left[ \sqrt{{\tilde J}} 
\ln (D_l/T) 
\right] 
& \, \mbox{I} \\
\frac{\displaystyle{
\sqrt{{\tilde J}} 
\tan \left[ \sqrt{{\tilde J}} 
\ln (D_l/{\mit\Gamma} \kappa^3) 
\right]
}
}{
\displaystyle{
1- \sqrt{{\tilde J}} 
\tan \left[ \sqrt{{\tilde J}} 
\ln (D_l/{\mit\Gamma} \kappa^3) 
\right]
\ln ({\mit\Gamma} \kappa^3/T)
}
} 
&  \mbox{II}
\end{array}\right. ,
\]

\noindent 
where ${\tilde J}$$=$$(J \varrho $$/$$6N)(\kappa_0^2$$/$$ p_{\rm F}^2)$. 
Since $\lambda_0(0)$$\geq$$\lambda_l(0)$ for arbitrary $l$, 
we evaluate $T_{\rm K}$ as temperature at which 
$\lambda_0(0)$ diverges. 
Noting ${\mit\Gamma} \kappa^3$$\simeq$$D_0 r^{3/2}$ 
in the region II, we obtain 

\[
\frac{T_{\rm K}}{D_0} = \left\{
\begin{array}{ll}
\exp \left( - \pi \big/ 2 \sqrt{{\tilde J}} \right) 
&  \quad r<r^* \\
r^{3/2}
\exp \left( - 1 \big/ 
\sqrt{{\tilde J}} 
\tan \left[ \sqrt{{\tilde J}} | \ln r^{3/2} | \right] \right) 
&  \quad r>r^*
\end{array}\right. ,
\]

\noindent
where $r^*$$\equiv$$\exp \left( - \pi / 3 \sqrt{{\tilde J}} \right)$. 
As a direct consequence of the coupling between the quantum fluctuation of 
an individual magnetic impurity and dynamical spin fluctuations of host 
electrons, $T_{\rm K}$ is seen to be enhanced in the region around 
a ferromagnetic QCP.

On the basis of the two-loop expansion, 
the quasi-particle's damping rate 
$\tau_{\rm imp}^{-1} (r,T)$$=$$ z {\rm Im} \Sigma_{\rm imp} (r,T)$ and 
the electrical resistivity $\rho_{\rm imp} (r,T)$ 
due to magnetic impurities can be 
expressed in terms of $\lambda_l(\omega)$ as follows:

\begin{eqnarray}
& & \tau_{\rm imp}^{-1} 
=  \frac{\pi^2}{4} S(S+1) 
\sum_{l=0}^{\infty} (l + 1)^2 (\lambda_l^2(0)- \lambda_{l+1}^2(0)),
\label{Eq13}
\\
& & \rho_{\rm imp} 
= \frac{\pi^2}{4} S(S+1) 
\sum_{l=0}^{\infty} (l+1) (\lambda_l(0) - \lambda_{l+1}(0))^2.
\label{Eq14}
\end{eqnarray}

\noindent
By solving Eq.(\ref{beta}) numerically 
by use of Eq.(\ref{ReGamma}) for $\lambda_l^{({\rm b})}(\omega)$, 
$T$-dependences of $\tau_{\rm imp}^{-1}$ and $\rho_{\rm imp}$ 
are obtained in Fig.2, 
for ${\tilde J}$$=$$0.02$ in the weak coupling regime 
and for several values of $r$.
We also show the result of the Born approximation 
where $\tau_{\rm imp}^{-1}$$\propto$$\xi^2(T)$ and 
$\rho_{\rm imp}$$\propto$$\ln \xi(T)$, 
which visualizes the crossover line from the regime I to the regime II.  
The fixed point is that of 
the single-channel Kondo problem; 
the multichannel effects 
can be seen as a transient phenomenon. 
Note that $T_{\rm K}$, 
at which $\rho_{\rm imp}$ is of the order of $1$, 
increases rapidly in the region II 
from $r$$\sim$$0.1$ to $\sim$$0.01$ 
while showing a tendency to be saturated in the region I, 
consistently with the one-loop result. 

In the regime II far away from the QCP, 
the $\omega$-dependence of the vertex $\Gamma_l^{(\sigma)}$ is suppressed, 
so that the low-energy effective Hamiltonian may be mapped into the 
anisotropic $n$$\approx$$(\l_{\rm max}$$+$$1)^2$-channel Kondo model 
with the band-width cutoff $D$$\approx$$D_0$$r^{3/2}$, 
in which the coupling constant $\lambda_l$ is given by 
$\lambda_l^{({\rm b})} (0)$.
By general considerations of the anisotropic multichannel Kondo model, 
while one enters the influence of the multichannel fixed point 
for $T$ below $O(T_{\rm K})$, 
channel anisotropy introduces another energy scale $T_{\rm x}$ 
below which the stable single-channel fixed point is reached\cite{ALPC92}. 
Hence the crossover from the multichannel to the single-channel behavior 
is predicted when $T_{\rm x}/T_{\rm K}$$\ll$$1$.
For the $n$-channel case, 
$T_{\rm K}$$\sim$$D$${\bar \lambda}^{n/2} \exp (-1/{\bar \lambda})$ 
and $T_{\rm x}$ asymptotically scales as $(\Delta$$\lambda)^{1+n/2}$, 
where ${\bar \lambda}$ and $\Delta$$\lambda$ are the average and difference 
of coupling constants, respectively\cite{ALPC92}. 
For the two-channel case, a recent numerical RG study 
has shown that $T_{\rm x}$$\simeq$$D$$(\Delta$$\lambda$$/$${\bar \lambda})^2 
\exp (-1/{\bar \lambda})$\cite{YMOO}. 
From these results, 
we expect that $T_{\rm x}$ is given by 
$D$$(\Delta$$\lambda$$/$${\bar \lambda})^{1+n/2} \exp (-1/{\bar \lambda})$, 
so that $T_{\rm x}/T_{\rm K}$ can be evaluated as $(\Delta 
\lambda)^{1+n/2}$$/$${\bar \lambda}^{1+n}$. 
For the present effective Hamiltonian, 
from Eq.(\ref{approx}), 
$\lambda_l$$\simeq$$3 {\tilde J} |\ln \sqrt{r}|$ 
while $\lambda_l$$-$$\lambda_{l+1}$$\simeq$$3 {\tilde J}$$/$$(l$$+$$1)$. 
Since $l_{\rm max}$$\approx$ $r^{-1/2}$, 
we may make a rough estimation of $T_{\rm x}$$/$$T_{\rm K}$ as 

\begin{equation}
\frac{T_{\rm x}}{T_{\rm K}} \sim 
(3 {\tilde J})^{-n/2} |\ln \sqrt{r}|^{-n-1}
\approx
\left( \sqrt{3 {\tilde J}}|\ln \sqrt{r}| \right)^{-1/r}.
\end{equation}
At $r$$=$$0.1$ this is $O(10^{-3})$ and at $r$$=$$0.05$ 
of $O(10^{-8})$ for ${\tilde J}$$=$$1$. 
Therefore there is a possibility of $T_{\rm x}$$\ll$$T_{\rm K}$ 
in the intermediate coupling regime. 
Then for the region of interest away from the QCP in the region II, 
at experimentally attainable temperatures the 
observable properties are expected to be 
those for a line of critical points with singular properties 
with exponents continuously changing as $r$ changes.

The true fixed point at $r$$\geq$$0$ would be the fixed point of 
the single-channel Kondo problem. It has zero ground state entropy. 
For finite $r$ and $T$$\gg$$T_{\rm x}$ in the intermediate coupling regime,  
we may read off the results for the entropy and its leading 
temperature dependence
from the exact solution for the multichannel problem \cite{AL91-1}.

On the experimental side, systematic results for the resistivity 
as a function of $r$ are not available. 
Further work is required to test the predictions 
made above.

Authors would like to acknowledge H. Kohno, O. Narikiyo, T. Senthil 
and Qimiao Si for clarifying discussions. 
They also thank I. Affleck, D. L. Cox, H. Kusunose and S. Yotsuhashi 
for useful comments on the multichannel Kondo problem. 
H. M. was supported by the Japan Society for the Promotion 
of Science for Young Scientists. 
He also thanks Bell labs and Lorentz Center 
for hospitality during his visits. 
This work is partly supported by a Grant-in-Aid 
for COE research (10CE2004) by the Ministry of Education, 
Science, Sports and Culture.

 \begin{minipage}{8.2cm}
 \begin{figure}  
 \begin{center}
 \epsfxsize=7.2cm  \epsfbox{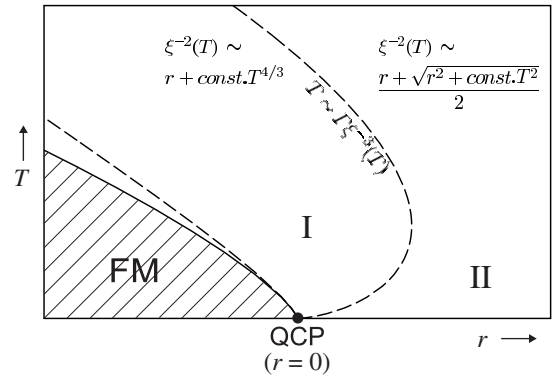}
 \end{center}
 \caption{A schematic phase diagram near a Ferromagnetic
 Quantum Critical Point.}
 \label{}\end{figure}
 \end{minipage}
 
 \begin{minipage}{8.2cm}
 \begin{figure}  
 \begin{center}
 \epsfxsize=7.2cm  \epsfbox{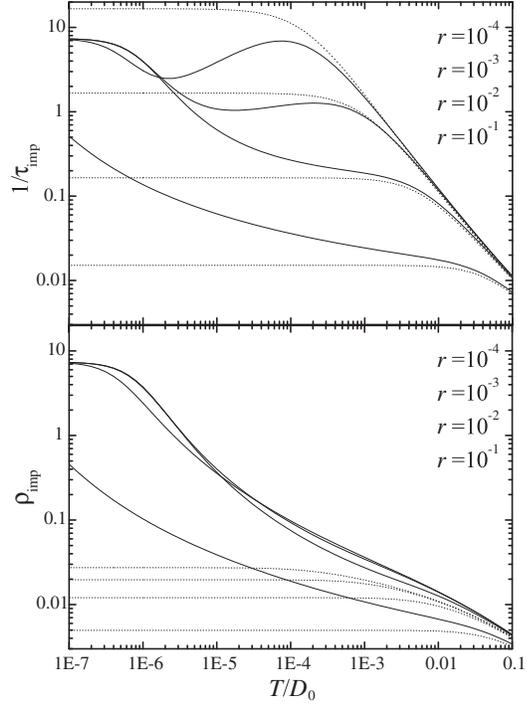}
 \end{center}
 \caption{$T$ dependences of $\tau^{-1}_{\rm imp}$ and $\rho_{\rm imp}$ 
 for ${\tilde J}$$=$$0.02$, 
 at different $r$'s ($r$$=$$10^{-4}$,$10^{-3}$,$10^{-2}$, and $10^{-1}$ 
 going down, starting from the top curves at high $T$) 
 on the basis of the two-loop expansion (solid lines) and 
 the Born approximation (dotted lines). }
 \label{}\end{figure}
 \end{minipage}

\end{multicols}
\end{document}